# A "one-size-fits-most" walking recognition method for smartphones, smartwatches, and wearable accelerometers

Marcin Straczkiewicz[1]*, Emily J. Huang[2], Jukka-Pekka Onnela[1]

[1] Department of Biostatistics, Harvard University, Boston, MA 02115, USA

[2] Department of Statistical Sciences, Wake Forest University, Winston Salem, NC 27106, USA

* Corresponding author

## Abstract

The ubiquity of personal digital devices offers unprecedented opportunities to study human behavior. Current state-of-the-art methods quantify physical activity using "activity counts," a measure which overlooks specific types of physical activities. We proposed a walking recognition method for sub-second tri-axial accelerometer data, in which activity classification is based on the inherent features of walking: intensity, periodicity, and duration. We validated our method against 20 publicly available, annotated datasets on walking activity data collected at various body locations (thigh, waist, chest, arm, wrist). We demonstrated that our method can estimate walking periods with high sensitivity and specificity: average sensitivity ranged between 0.92 and 0.97 across various body locations, and average specificity for common daily activities was typically above 0.95. We also assessed the method's algorithmic fairness to demographic and anthropometric variables and measurement contexts (body location, environment). Finally, we have released our method as open-source software in MATLAB and Python.

The development of body-worn devices, such as smartphones, smartwatches, and wearable accelerometers, has revolutionized research on physical activity (PA) in medicine and public health. Unlike surveys, which are subjective and often cross-sectional, body-worn sensors collect objective and continuous data on human behavior. The personal nature of body-worn sensors and their ability to collect high-resolution data allow researchers to obtain insights into everyday activities, thus deepening our understanding of how PA impacts human health.

Human activity recognition (HAR) is the process of translating discrete measurements from body-worn devices into physical human activities that may occur in the lab or in free-living settings [1]. In public health research, body-worn devices can be used to quantify PA in terms of "activity counts," which classify activities based on their intensity level (traditionally expressed in gravitational units [g] [2]) using predefined thresholds developed for each body location where the sensor is carried [3–5]. PA in a given period of observation may be classified as sedentary, light, moderate, or vigorous. One drawback of classifying PA by intensity is that it overlooks the importance of specific types of activities, which depend on personal capabilities, choices, habitual changes, and detailed characteristics of motion, which could indicate deteriorating health status. As a potential alternative, human activities may be classified by type, replacing PA intensity levels with the type of activity performed, e.g., walking or running. Such an approach requires an understanding of how different activities manifest themselves as measurable physiological motion.

In this study, we focused on the recognition of walking using a wide spectrum of personal digital devices, such as smartphones, smartwatches, or wearable accelerometers. Walking is the most common PA performed daily by able-bodied humans starting approximately from the age of one year [6]. Walking not only allows us to commute, but also serves as an essential exercise that helps to maintain healthy body weight and prevent disease, for example, heart disease, high blood pressure, cognitive decline, and type 2 diabetes [7–11]. The increasing application of body-worn devices in free-living epidemiological studies is expected to provide new insights into quality of life [12], as well as allow exploration and possible extension of walking-related biomarkers, such as cadence, step length, and gait variability [13–15] across heterogeneous cohorts of subjects.

Walking recognition using body-worn devices is a challenging task, and it has not been implemented on a large scale using open (non-proprietary) methods (Supplementary Table 1). Walking measurements from body-worn sensors are complex and depend on not only demographic (e.g., age, sex), anthropometric (e.g., height, weight), and habitual (e.g., posture, gait, walking speed) differences among subjects, but also on metrological (e.g., sensor body location and orientation, body attachment, sensing device, environmental context) differences across studies. Figure 1a illustrates the variety of signals, such as walking strides (i.e., motion between two consecutive steps of walking), from several publicly available datasets that we used in our study (Table 1). The data were collected with accelerometers situated in various body-worn devices at different locations. To simplify comparison, we rescaled each walking fragment to the same length.

The data collected at a given body location within a given study exhibit visual similarities between subjects in terms of signal amplitude and variability; however, when compared across studies, walking signals are much more heterogenous. Despite some common features, such as a certain minimum amplitude and oscillations, the data representing the same activity exhibit different characteristics not just between body locations but, more importantly, within the same location. Since each dataset was collected in a different environment using different instrumentation and different data acquisition parameters, it is unclear whether existing methods can be adapted to these settings without compromising their classification accuracy [16,17]. Consequently, while existing methods offer solutions that are "fit-for-purpose," e.g., methods that have been developed for a specific cohort, device, and body location, the literature still lacks a "one-size-fits-all" or at least a "one-size-fits-most" method that provides accurate, generalizable, and reproducible walking recognition in various measurement scenarios, is insensitive to other everyday activities, and, importantly, is not systematically biased towards one specific group of subjects either in terms of demographic or anthropometric measurements.

Here, we proposed a method that recognizes walking activity through temporal dynamics of human motions measured by the accelerometer, a standard hardware sensor built in body-worn devices. Our approach focuses on the inherent features of walking: intensity, periodicity, and duration. We analyzed these features for sensors at body locations typically used in medical and public health studies (thigh, waist, chest, arm, wrist) as well as for unspecified locations (e.g., in free-living settings using smartphones), and created a classification scheme that allows for flexible and interpretable estimation of walking periods and their temporal cadence. To account for diversity in walking, we validated our method against 20 publicly available datasets (Table 1). To assess the algorithmic fairness of our method, we evaluated our approach for a potential bias toward subjects' demographics and measurement context. To improve transparency and reproducibility of research, we have released open-source software implementations of our method in MATLAB and Python [18].

## Results

Our method leverages the observation that, regardless of sensor location and subject, as long as a person is walking, their accelerometer signal oscillates around a local mean with a specific amplitude and a frequency equal to their walking speed (Figure 1). To determine signal amplitude, we computed a peak-to-peak distance in one-second non-overlapping segments; information about temporal characteristics was obtained using continuous wavelet transform (CWT) (Figure 2). The algorithm is discussed in full detail in the Methods section.

Our method has several tuning parameters. To account for substantial differences in frequency-domain features across body locations (Figure 3), we optimized our algorithm for two possible application scenarios: (1) smartphone or waist-worn accelerometer data (i.e., when device is typically carried on thigh, waist, chest, or arm); and (2) smartwatch and wrist-worn accelerometer data (i.e., device is typically carried on the wrist).

Method evaluation was performed using data from 1240 subjects in 20 publicly available datasets (Table 1). Cumulatively, our analysis included more than 831 h of accelerometer measurements split into 56,467 bouts, with more than 267 h of data representing various types of walking, such as flat walking, climbing stairs, or walking on a treadmill (15,234 bouts) collected at various body locations: thigh – 55 h (2593 bouts); waist – 69 h (2460 bouts); chest – 11 h (544 bouts); arm – 9 h (197 bouts); and wrist – 67 h (1829 bouts); and 54 h (7611 bouts) collected at unspecified locations.

ROC curves (Figure 3c) were used to select optimal thresholds for tuning parameters. Thresholds for $A$ and $f_w$ were similar for the smartphone and smartwatch, and were set to $\hat{A} = 0.3$g, and $\widehat{f_w} = [1.4 \text{ Hz}, 2.3 \text{ Hz}]$ (values rounded to two significant figures). Thresholds for $\alpha$, $\beta$, and $T$ differed between the two devices and were set to $\hat{\alpha} = 0.6$, $\hat{\beta} = 2.5$, and $\hat{T} = 3$ for the smartphone and $\hat{\alpha} = 31.7$, $\hat{\beta} = 1.4$, and $\hat{T} = 6$ for the smartwatch. These choices resulted in $AUC_A = 0.848, AUC_{f_w} = 0.959, AUC_{\alpha,\beta} = 0.965,$ and $AUC_T = 0.961$ for smartphones and $AUC_A = $

0.850, $AUC_{f_w}$ = 0.954, $AUC_{\alpha,\beta}$ = 0.968, and $AUC_T$ = 0.959 for smartwatches, indicating very good performance.

The estimated classification accuracy metrics (Table 2) suggest very high sensitivity (ranging between 0.92 and 0.97) for normal walking and across various sensor body locations. Sensitivity was somewhat lower for ascending stairs (min: 0.73, max: 0.93); descending stairs (min: 0.73, max: 0.86); and other variants of walking (min: 0.47, max: 0.81). The algorithm underperformed during slow treadmill walking at 1 mph (min: 0, max: 0.19), most likely due to very low gait speed, which is atypical in normal walking. Compared to other sensor locations, a very low sensitivity was noted at the wrist for a 2 mph walk (0.05, 95% CI: 0.01, 0.09), which might be due to rail holding that effectively damped motion acceleration.

The results also suggest that our method does not overestimate walking during most everyday activities. In the cases of sedentary periods, such as during desk work, eating, drinking, using motorized transportation, running, and cycling, the mean specificity scores are predominantly above 0.95 with a marginally better performance at locations typical to the smartphone. More profound dissonance was noted for selected household activities, e.g., the estimated specificity for sweeping was 0.94 (95% CI: 0.91, 0.97) for the smartphone, compared to only 0.57 (95% CI: 0.51, 0.62) for the smartwatch, likely due to the repetitive hand movements involved in sweeping. Regardless of sensor placement, specificity was systematically low for jumping, as this activity produces high acceleration with periodicity similar to normal walking.

Visual investigation of normal walking sensitivity scores indicated no systematic bias for any investigated demographic or body measure covariate (Figure 3d). At the aggregate level, the greatest difference in weight corresponded to a change of 0.02 in sensitivity (0.98 for 85 kg vs. 1.00 for 141 kg), 0.01 for height (0.98 for 1.70 m vs. 0.99 for 1.96 m), 0.01 for BMI (0.97 for 27 kg/m² vs. 0.98 for 15 kg/m²), and 0.02 for age (0.97 for 22 y vs. 0.99 for 24 y). These differences were greater at the level of individual datasets: 0.16 for weight (0.72 at 70 kg and 0.88 for 82 kg) in *UniMiBSHAR*, 0.21 for height (0.75 for 1.83 m vs. 0.96 for 1.63 m) in *SisFall*, 0.12 for BMI (0.88 for 25 kg/m² vs. 1.00 for 19 kg/m²) in *Actitracker*, and 0.17 for age (0.70 for 21 y vs. 0.87 for 29 y) in *UniMiBSHAR*.

We used a linear mixed-effects regression model to assess the effect of certain covariates on the algorithm's sensitivity score for normal walking, defined as the proportion of correct classifications of normal walking for a given sensor location. If a subject was tested with more than one sensor location, a separate sensitivity score was calculated for each. The covariates of interest were included as fixed effects, and the model also contained a random intercept for the subject. The random effect was included to account for the fact that some participants contributed multiple sensitivity scores (corresponding to different locations), and we expected the scores from the same participant to be correlated. The linear mixed-effects regression is referred to as *MixedReg*. We also performed the regression without the random effect (i.e., a standard linear regression), hereafter referred to as *StandardReg*.

Table 3a shows the estimates, standard errors, and confidence intervals for *StandardReg*. The column shows the covariates, including age, gender, BMI, sensor location (arm, chest, thigh, wrist, waist, or unspecified), environmental condition (controlled or free-living), and study (e.g., *Actitracker*, *DaLiAc*). Based on the 95% confidence intervals, we found that several covariates, including certain sensor locations and studies, were statistically significant using Type 1 error rate of $\alpha = 0.05$. To understand the influence of different studies, more information about the study settings would be required. Two specific sensor locations, chest and waist, were also statistically significant. The higher sensitivity scores for chest and waist likely result from the fact that the accelerometer can be more firmly attached to the body at these sites. The coefficients for age, gender, BMI, and environmental condition were not statistically significant.

In *MixedReg* (Table 3b), the coefficients for gender, BMI, and environmental condition were also not statistically significant. Furthermore, *MixedReg* had the same statistical significances for studies as *StandardReg*. On the other hand, *MixedReg* showed somewhat different results than *StandardReg* for sensor location and age. The coefficient estimates for chest and waist were closer to 0, and these fixed effects were not statistically significant in *MixedReg*. This result may be more reliable than that from *StandardReg* because *MixedReg* accounts for the nested structure in the data. In *MixedReg*, the coefficient for age was statistically significant, unlike for *StandardReg*. The *MixedReg* results suggest that older age is associated with higher sensitivity score. This difference from *StandardReg* may be related to the fact that, in our dataset, people of older ages accounted for a smaller portion of the total subjects. Also, older subjects were slightly more likely than younger subjects to contribute multiple observations. Overall, including the random effect in *MixedReg* indicates a stronger effect of age.

## Discussion

The application of body-worn devices in health studies allows for objective quantification of human activity. The domain, however, suffers from a lack of widely validated methods that provide efficient, accurate, and interpretable recognition of detailed PA types, such as walking. This gap is likely related to the heterogeneity of walking, which is substantially affected by several factors such as age, sex, walking speed, footwear, and walking surface; sensor data on walking is affected foremost by sensor body location (Figure 1a). For this reason, many studies have adopted approaches based on PA intensity levels, and various activity recognition methods have been developed for specific sensor body locations and specific populations. The methods in the literature have been predominantly validated using (1) a limited number of datasets that include small cohorts of subjects recruited from a specific population, e.g., college students or elder adults (Table 1), and (2) a limited number of body locations, often representing a subset of locations where the device might be carried in a real-life setting (especially for smartphones). In addition, (3) classification methods have been mainly trained and tested using specific measurement settings, sensor body locations and, occasionally, device orientations. These steps, which are aimed at simplifying the problem, appear to be either insufficient for describing real-life scenarios [12,19] or impractical to implement [20].

In this paper, we describe a method intended to fill this gap in the literature. Our method is based on the observation that regardless of sensor location, subject, or measurement environment, walking can be captured using body-worn accelerometers as a continuous and periodic oscillation with quasi-stationary amplitude and speed. We applied our method to data from 1240 subjects gathered in 20 publicly available datasets, which provide a large variety of walking signals and other types of PA. Our classification scheme makes use of signal amplitude, walking speed, and activity duration, i.e., features that are activity-specific rather than location- or subject-specific. The validation of our approach showed very good classification accuracy for normal walking, and good classification accuracy of other types of walking, e.g., stair climbing (Table 2). Notably, the method's performance is not sensitive to various demographic and metrological factors for individual subjects (Table 3).

The validation was conducted for each body location separately. Given that different devices are carried differently, we conclude that our method performs well when applied to a smartphone or wearable accelerometer placed at the waist, on the chest, or on the lower back, while method performance may be lower in real-life settings that employ a smartwatch or wrist-worn accelerometers due to vigorous and repetitive hand movements, such as those during household activities. Importantly, our method does not overestimate walking in the presence of other daily activities, such as sitting or driving, or for repetitive activities, such as running or cycling.

Our method was designed with two goals in mind: (1) robustness to heterogeneous devices and (2) computational performance. The first goal was achieved by employing only one sensor, an accelerometer, and limiting the required sampling frequency to 10 Hz. Accelerometers have become a standard tool for assessment of PA, and although recent technological advances have allowed researchers to benefit from ever more "sensored" devices, many ongoing health studies still use only the accelerometer to measure PA [25,26]. With regard to the selection of sampling frequency, our main consideration was to prevent excessive battery drainage in smartphones and smartwatches. Even though these devices are capable of collecting accelerometer data at very high rates (100 Hz or higher), such high frequencies require frequent battery charging [21]. The sampling frequency of 10 Hz is supported by the vast majority of available smartphones and wearables and is expected to provide longer battery life for data collection [22].

Limiting sampling frequency also benefited our second goal of computational performance. Given that our method employs CWT, which has computational complexity of $\mathcal{O}(N \cdot \log(N))$, we aimed at a sampling frequency just low enough to capture all typical everyday activities that cannot be filtered out using basic time-domain features (e.g., running [23]). Our method was also made computationally efficient with the use of the amplitude threshold $A$, which not only excluded large chunks of sedentary activities, such as lying, sitting, doing office work, and driving, but also efficiently limited the size of the input to CWT. When run on a standard desktop computer using a single core, the total execution time of our code for one subject on a week-long dataset was between 10 s and 20 s (excluding data uploading), which is sufficient for large-scale studies.

There are some limitations to our method. First, our method tends to systematically overestimate the duration of walking periods during exertional activities, such as rope jumping, particularly due to their significant overlap with walking features in both time and frequency domains. More sophisticated methods are needed to address this issue with accelerometer data; for example, GPS data could be used to measure geospatial displacement of the device and exclude periods when a subject was not moving around. This solution, however, may be valid mainly in outdoor settings, since GPS has limited indoor reception [24,25]. Second, our method was validated only on healthy subjects. For reasons of reproducibility, we only considered publicly available datasets. More research is needed to determine walking characteristics in individuals who have walking impairments or use walking aids, such as canes or walkers. Third, our method was validated on a limited number of elder adults, and it was not validated on children. Given that these groups might walk differently than the investigated population [26,27], our method needs to be used with caution and changing the amplitude threshold and step frequency range may be required. In this case, a potential overlap with activities that contain low-amplitude low-frequency vibrations, such as car driving, might be addressed using dedicated methods [28,29]. Fourth, our investigation included only four datasets collected in free-living conditions, and in two of these (*Actitracker* and *Extrasensory*), the activities were labelled by the study participants. The labelling in these datasets suffered visual discrepancies, and although we tried to correct labels in the most prominent cases (e.g., when period of flat acceleration was labelled as walking), the accuracy metrics estimated at *unspecified* locations (Table 3) are not fully representative of our method and need further investigation.

In summary, we proposed a method for walking recognition using various body-worn devices, including smartphone, smartwatch, and wearable accelerometers. A robust validation demonstrated that our approach adapts to various walking styles, sensor body locations, and measurement settings, and it can be used to estimate walking time, cadence, and step count.

# Methods

## Acceleration signal of walking activity

Kinesiology describes walking as a cyclic series of movements initiated the moment the foot contacts the ground, followed by the stance phase (i.e., when the foot is on the ground) and the swing phase (i.e., when the foot is in the air); the cycle is completed when the same foot makes contact with the ground again [30] (Figure 1b). The fundamental challenge in walking recognition using accelerometer data from various body-worn devices results from the fact that these movements are reflected differently in data depending on several factors, including sensor location and subject. Figure 1c displays several examples of resting and walking acceleration signals collected using smartphones at different body locations (thigh, waist, chest, arm) and smartwatches worn on the wrist by two subjects. The univariate vector magnitude was determined by transforming the raw data from the three orthogonal vectors. These data were obtained from the publicly available HAR dataset called *RealWorld* [31]. According to the supplementary video recordings available for that study, the subjects wore sport shoes during data collection and performed activities on concrete pavement.

When a sensor is placed on the thigh, one cycle of walking consists of the following stages: the heel strikes the ground (event I) and is registered as a spike, the body decelerates during balancing in the stance phase (between events I and II), the opposite heel strikes the ground (event II) and is registered as a somewhat lower spike, and finally the body accelerates in the swing phase until the cycle is completed with the heel striking the ground again (event I). In contrast, when the sensor is placed closer to the center of body mass (i.e., at the waist, on the chest, around the arm), the amplitude of gait events appears to be more symmetrical and therefore it is difficult to distinguish them from one another. A more confusing scenario occurs for a sensor placed on the wrist: for subject 1, the signal resembles that obtained from the thigh, whereas for subject 2, the signal resembles that obtained from the waist, chest, and arm. An explanation for these discrepancies may be deduced from the videos, which show that during the walking activity, subject 1 held her hand close to the body, while subject 2 performed arm swings.

The complexity of walking recognition is magnified by the fact that each of the displayed fragments contains a different repetitive template of acceleration not only among body locations, but also across subjects. Moreover, the observations derived from Figure 1c might not replicate in different studies (e.g., see Figure 1a). What appears common to all investigated walking signals is the continuous and periodic oscillation of acceleration around a long-term average with quasi-stationary amplitude and speed. The panels corresponding to time-domain signals display their time-frequency representations (scalograms) estimated using wavelet transformation, which shows the relative weights of different frequencies over time with brighter colors indicating higher weights. Regardless of sensor location and subject, as long as the person is walking, the periodic components hover around 1.7 Hz, which corresponds to the published range of human walking speed between 1.4 Hz and 2.3 Hz (steps per second) [32,33]. Depending on sensor location and walking characteristics, the predominant *step* frequency may be accompanied by both subharmonics (resulting from a limb swing at half of step frequency, also called the *stride* frequency) and higher harmonics (resulting from the energy dispersion during heel strikes at multiples of the *stride* frequency) [34,35]. The subharmonics are therefore likely to appear on the wrist, as this location is prone to swinging during walking. On the other hand, the higher harmonics are likely to manifest closer to the lower limbs. The higher harmonics are also likely related to other factors, including demographics, style of walking, footwear, type of surface a person walks on, as well as sensor body attachment. In our approach, we leverage the common features of walking: quasi-stationary amplitude, specified gait speed, and activity duration.

**Continuous wavelet transform**

The time-frequency distributions presented in Figure 1c were obtained using a wavelet projection approach, which decomposes the original signal into various frequencies. Specifically, we used continuous wavelet transform (CWT) to capture the globally non-stationary but locally quasi-periodic characteristics of walking. Indeed, while one can assume that walking is quasi-periodic for a short period of time (e.g., the time between consecutive steps is roughly equal when a person walks along a hallway), walking characteristics can change dramatically over the course of a day due to the individual's level of energy, environmental context, and goals. CWT decomposes the original signal $v(t)$ into a set of scaled time-shifted versions of a prespecified 'mother' wavelet $\psi(t)$ using the transformation $C(f,\tau) = \frac{1}{\sqrt{|f|}} \int_{-\infty}^{+\infty} v(t) \cdot \psi\left(\frac{t-\tau}{f}\right) dt$, where $f$ is the frequency scale and $\tau$ is the time-shift. By

continuously scaling and shifting the mother wavelet, the original signal is projected onto the time-frequency space. The result of this transformation, wavelet coefficients, represent the similarity between a specific wavelet function, characterized by $f$ and $\tau$, and a localized section of the signal $v(t)$. Thus, wavelet coefficients are maximized when a particular frequency, $f$, matches the frequency of the observed signal at a particular time point. Because of this construction, CWT is sensitive to subtle changes, breakdown points, and signal discontinuities. This is essential in walking recognition, where both subtle and sudden changes in walking frequency are the norm. Moreover, unlike Fourier transform used in previous studies (Table 1), CWT does not depend on a particular window size and does not require a prespecified number of repetitions of the activity to estimate the local frequency.

**Walking recognition algorithm**

We let the measured signal be $x(t) = (x_1(t), x_2(t), x_3(t))$, where $x_1(t), x_2(t)$, and $x_3(t)$ denote the measurements along each of the orthogonal axes of the device at time $t$ in units of g. After the initial two preprocessing steps described below, in the section *Data preprocessing*, we transformed the signal to its vector magnitude form $v(t) = \sqrt{x_1(t)^2 + x_2(t)^2 + x_3(t)^2} - 1$ (Figure 2). We then estimated the periods when the sensor recorded intensive body motions. For this purpose, we split the signal into consecutive and non-overlapping one-second windows and calculated the peak-to-peak amplitude in each window. This metric was then compared with a threshold $A$. Segments with amplitude below the threshold were excluded from further consideration. In a typical scenario, consecutive steps occurred in intervals roughly between 0.43 s and 0.71 s for a walking speed between 2.3 steps per second and 1.4 steps per second, respectively. The one-second window length was selected to ascertain that during walking activity, there was at least one step-related spike in each consecutive time window.

In the next step, we computed CWT over the high-amplitude segments to obtain their projection onto the time-frequency domain $C(f, \tau)$. Specifically, we used the generalized Morse wavelet as the mother wavelet, defined as $\Psi_{P,\gamma}(\omega) = U(\omega) a_{P,\gamma} \omega^{\frac{P^2}{\gamma}} e^{-\omega^\gamma}$, where $U(\omega)$ is the unit step, $a_{P,\gamma}$ is a normalizing constant, $P^2$ is the time-bandwidth product, and $\gamma$ characterizes the symmetry of the Morse wavelet [36]. Here we used $\gamma = 3$ and $P^2 = 60$, which produced coefficients spread symmetrically both in time- and frequency-domains, i.e., skewness around the peak frequency was close to or equal to 0 in time and frequency domains, respectively [37,38] (Supplementary Figure 1). Other choices for mother wavelets for our method were the Morlet and Bump wavelets.

As depicted in Figure 1, while some walking signals might be represented by a series of harmonics, the information that was consistently preserved throughout, regardless of sensing device and walking pattern, was present within a certain step frequency range $f_w = [f_{min}, f_{max}]$, where $f_{min}$ and $f_{max}$ are statistically derived minimum and maximum frequencies, respectively. To account for this fact and the presence of harmonics, we created a new vector, $w(\tau)$:

$$w(\tau) = \begin{cases} 1 & \text{if} \\ 0 & \text{otherwise} \end{cases} \quad \alpha \cdot \max_{f \in f_w}(C(f,\tau)) > \max_{f < f_{min}}(C(f,\tau)) \land \beta \cdot \max_{f \in f_w}(C(f,\tau)) > \max_{f > f_{max}}(C(f,\tau))$$

Here, the parameters $\alpha$ and $\beta$ control the ratio between the maximum wavelet coefficients that fall below and above $f_w$, respectively, and allow for flexible accounting of harmonics related to, e.g., heel strikes or arm swings. They also prevent capturing other periodical activities with local maxima within $f_w$ which are sub- or higher harmonics of other processes with global maximum frequency outside of $f_w$ (e.g., *stride* frequency of running).

Finally, an activity was identified as walking when $w(\tau)$ is positive for $T$ consecutive windows. The selection of tuning parameters leads to a trade-off between sensitivity and specificity of walking classification accuracy in any given study. For instance, using a large $T$ (e.g., $T = 10$) will result in a higher specificity as fewer non-walking activities generate oscillations within $f_w$ that long, but it will also miss shorter walking bouts.

In the following sections, we discuss the selection of tuning parameters ($A, f_w, \alpha, \beta,$ and $T$) based on the walking characteristics extracted from several publicly available studies.

## Data description

To validate our method, we identified 20 publicly available datasets with at least 10 subjects each that contain accelerometer data from smartphones, smartwatches, or wearable accelerometers along with activity labels on various types of PA (Table 1). Walking activity was recorded in 19 studies, in all but *SpeedBreaker*. The datasets were collected by independent research groups in several countries worldwide, including the Netherlands, Italy,

Germany, Spain, Greece, Turkey, Colombia, India, Japan, and the United States. The aggregated dataset includes measurements collected on 1240 healthy subjects. Sex was provided for 901 subjects (649 males), age was provided for 745 subjects (between 15 and 75 years of age, mean±SD=28.6±12.0), height was provided for 865 subjects (147-196 cm, 170.6±8.6), and weight was provided for 858 subjects (37-141 kg, 66.2±14.2). Given available information, we calculated BMI for 858 subjects (15.1-39.8 kg·m$^{-2}$, 22.6±3.8). Cumulatively, a complete set of sex, age, height, weight, and BMI was available for 725 (58%) subjects.

Importantly, the datasets were collected under various measurement conditions, with different study settings (controlled, free-living), environmental contexts (indoor, outdoor), sensing devices (smartphones, smartwatches, data acquisition parameters), and body attachments (loose in pocket, affixed with a strap), which introduces considerable signal heterogeneity that is essential in validating any HAR algorithm aimed for real-life settings [1]. A summary of the investigated datasets is provided in Table 1.

Accelerometer data were collected using various wearable devices, primarily smartphones (running the iOS or Android operating system) and smartwatches of various manufacturers; a few studies used research-grade data acquisition units, such as various versions of SHIMMER (Dublin, Ireland) and ActiGraph (Pensacola, Florida), or devices developed by the research groups themselves. The devices were positioned at various locations across the body. In our study, we focused on measurements collected at body locations typical to the devices' everyday use, i.e., around the thigh, at the waist, on the chest, around the arm, and on the wrist. We also analyzed measurements taken when the device location was unspecified. For example, in *SpeedBreaker*, the researchers randomly placed the smartphone in the pants pocket, cupholder, or below the windshield, while in *Actitracker*, *Extrasensory,* and *HASC*, smartphones were placed according to the subjects' preferences.

As the devices were selected and placed independently by each research group, and the devices' exact location and orientation differed between studies. This closely mimics a real-life situation when a researcher is confronted with a dataset from a subject who carried the device according to his or her individual preferences [39]. In our study, we grouped measurements from devices placed in similar locations into categories. For example, if the device was carried in the pants pocket, we treated it as being on the thigh. If it was carried on a waist belt, on the hip, or on the lower back, we treated it as being on the waist. If it was carried in a shirt or jacket pocket, or strapped around one's chest, we treated it as being on the chest. If it was carried in hand or on a forearm, we treated it as being on the wrist.

Measurement parameters also differed across the devices. The studies reported sampling frequencies between 20 Hz and 205 Hz. In some studies, the actual sampling frequency deviated from the requested one by a few to several Hz. The reported measurement range was between ±1.5 g and ±16 g (very high values of acceleration arose in studies that investigated falls), while the amplitude resolution (bit depth) was between 6 bit and 13 bit.

The participants performed a wide range of PA types. Depending on the study scope and aim, the performed activities included various types of walking, leisure activities, motorized transportation, household activities, recreational sports, etc. In *Extrasensory, HAPT, HASC, HMPD, MobiAct, MotionSense, SFDLA, SisFall,* and *UniMiBSHAR*, activities were recorded in several trials. Activity labelling was carried out in one of two ways: (1) in studies conducted under controlled conditions, activity labels were recorded by trained researchers, whereas (2) in free-living settings, labeling was performed either by researchers (*HASC, SpeedBreaker*) or by study participants using dedicated smartphone applications (*Actitracker, Extrasensory*). In a few studies (*Actitracker, MHEALTH, SisFall, Pedometer,* and *WISDM*), the investigated activities also included various falls, stumbles, or complex activities. These activities might have contained intermittent periods of walking; however, we excluded them from consideration due to the lack of precise timing of walking start and end. Additionally, we did not analyze data collected when the device was not carried on the subject's body (*Extrasensory*).

We grouped certain similar activities in common categories: activities described as jogging or running were analyzed as running; self-paced flat walking, slow flat walking, and fast flat walking were considered as normal walking; forward and backward jumping, rope jumping, and jumping in place were analyzed as jumping, etc. A complete summary of activity groupings is provided in Supplementary Table 2.

**Data preprocessing**

We carried out a few data preprocessing steps to standardize the input to our algorithm. First, we verified if the acceleration data were provided in gravitational units (g); data provided in SI units were converted using the standard definition 1g=9.80665m/s$^2$. Second, we used linear interpolation to impose a uniform sampling frequency of 10 Hz across tri-axial accelerometer data. Third, to alleviate potential deviations and translations of the measurement device, we transformed the tri-axial accelerometer signals into a univariate vector magnitude, as described above in the section *Walking recognition algorithm*.

Visual investigation of the datasets revealed that, in several studies, the walking activity was preceded and succeeded by stationary activities (e.g., standing still) that manifested as flatlined accelerometer readings; however, the corresponding activity labels marked the entire activity fragment as walking. To address this issue, we adjusted walking labels to periods when the moving standard deviation, computed in one-second non-overlapping windows, was above 0.1 g for at least two out of three axes, practically limiting labelled walking to periods when there was any motion recorded.

**Tuning parameter selection**

Our method requires several input parameters, namely minimum amplitude $A$, step frequency range $f_w$, harmonic ratios $\alpha$ and $\beta$, and minimum walking duration $T$. To learn how these features reflect across walking data from various studies, we selected signals of normal walking and preprocessed these signals using the methods described above, in the section *Data preprocessing*. Vector magnitudes were then segmented into non-overlapping one-second segments, and we processed each segment using several statistical and signal processing methods described below. The extracted information was then accumulated within subject and visualized using heatmaps (Figure 3a) where each row corresponds to a subject while color intensity corresponds to the frequency of a given value for this subject. To allow visual comparison between subjects, the values were normalized to [0,1] intervals.

A peak-to-peak amplitude was calculated to determine typical walking intensity levels. This analysis revealed that the recorded walking signals spanned across a wide range of amplitudes ranging from about 0.4 g to 2.5 g (and most typically between 0.5 g and 1.5 g) and they were visually greater for sensors at lower body locations (thigh, waist) compared to upper body locations (chest, arm, wrist).

Computation of a CWT over the segmented walking signal revealed that the predominant step frequency ranged between 1.7 Hz and 2.2 Hz; in some studies (e.g., *IWSCD*), the speed span was slightly wider, e.g., between 1.4 Hz and 2.3 Hz. Even though this dataset mostly consisted of young adults observed in controlled settings (Table 1), we hypothesize that, in free-living settings, lower walking speed might be more common to elders, while higher walking speed might be more common to adolescents and children.

The wavelet coefficients showed that the step frequency is often accompanied by its sub- and higher harmonics. The sub-harmonics were predominantly present at the wrist, while higher harmonics were predominantly present at lower body parts, particularly the thigh (and impacts harmonic ratio $\alpha$.) As pointed out earlier, the appearance of sub-harmonics results from limb swings, while the appearance of higher harmonics is due to distortions of walking signal during stepping, which are naturally better damped at locations closer to the body torso. We also observed that the presence of harmonics is somewhat study-specific (e.g., compare *DaLiAc* and *WISDM* at wrist), which might be due to the different surfaces walked upon. Unfortunately, study protocols did not provide sufficient details to explore this phenomenon further. However, in contrast to lower body parts, the strong presence of sub-harmonics at the wrist suggests that, at this location, the acceleration resulting from steps might be considerably overshadowed by acceleration resulting from vigorous hand swings. This discrepancy between smartphone and smartwatch locations suggests that our method will perform better if supplemented with *a priori* knowledge about the sensing device, i.e., smartphone or smartwatch.

Walking duration depends on several factors, including individual capabilities, choices, and needs. Generally, walking is considered a series of repetitive leg movements (see above section, *Acceleration signal of walking activity*), but it is not clear how many of these repetitions are required to call the activity *walking*, i.e., whether it is one step, one stride, or multiple strides. This information is also not specified in the available datasets or referenced HAR methods. The smallest window size considered in our method is equal to 1 s, which corresponds to an approximate duration of one stride. However, walking recognition at that resolution might come with a decreased specificity due to the temporal similarity between motions performed during walking and during other everyday activities (e.g., hand manipulation during washing dishes captured at the wrist or body swinging during floor sweeping captured on the thigh). An improved classification specificity may be achieved using multiple windows aimed at recognition of walking bouts that consist of at least a few strides.

In the main evaluation, the optimal tuning parameters were selected using receiver-operating characteristic (ROC) curves in a one vs. all scenario where we compared normal walking with all non-walking activities. The calculations were carried out separately for body locations typical to smartphone and smartwatch. The area under the subsequent ROC curves (AUC) was used to estimate the quality of our algorithm at each step of activity classification. The optimal cutoff points for $A$, $f_w$, $\alpha$, $\beta$, and $T$ were defined as points at which the sum of sensitivity and specificity was maximized. The thresholds were then used to calculate walking recognition accuracy metrics and to assess bias toward cohort demographics and body measures.

**Method evaluation**

We evaluated the proposed method for the accuracy of walking recognition. First, we identified walking periods in PA measurements from the aggregated datasets. The outcome of the algorithm was compared with the provided activity labels. The accuracy was estimated using sensitivity (true positive rate) and specificity (true negative rate). Sensitivity was used to estimate classification accuracy for measurements that contained various walking activities (normal walking, ascending stairs, descending stairs, walking backward, treadmill walking), and was calculated as the ratio between the number of true positives and the sum of true positives and false negatives. Specificity was calculated for signals that contained other activities, and was calculated as the ratio between the number of true negatives and the sum of true negatives and false positives. If a subject performed multiple trials of a given activity, their scores were averaged. The resulting metrics were then averaged across all subjects performing a given activity and reported as mean and 95% confidence intervals (95% CI).

**Bias estimation**

We sought to determine whether the accuracy of our algorithm is influenced by certain subject characteristics or data collection settings. To address this question, a standard linear regression analysis was first performed, referred to as *StandardReg*. The response variable ($Y$) was a subject's sensitivity score for normal walking at a particular sensor location. The covariates in the model included a subject's age, gender, and BMI, as well as sensor location, environmental condition, and the study to which a subject belonged. The model equation for *StandardReg* was:

$$Y_{ij} = \beta_0 + \beta X_{ij} + \epsilon_{ij},$$

where $Y_{ij}$ is the sensitivity score for subject $i$ at sensor location $j$, $X_{ij}$ is the vector of covariates, $\beta_0$ is the y-intercept, $\beta$ is the vector of coefficients for the covariates, and $\epsilon_{ij}$ is random noise. We then performed a separate linear mixed-effects regression analysis (*MixedReg*) to account for clustering in the data. The model equation for *MixedReg* was:

$$Y_{ij} = \beta_0 + \beta X_{ij} + b_i + \epsilon_{ij}.$$

The model equation is similar to that of *StandardReg*, except that *MixedReg* incorporated a random intercept ($b_i$) for each subject $i$, called a random effect.

In both analyses, we calculated 95% confidence intervals to assess statistical significance of the coefficients in the vector $\beta$. Conventional confidence interval formulas based on t values were used for *StandardReg*, and the percentile bootstrap was used for *MixedReg*. Since some subjects had missing values for certain covariates (age, gender, or BMI), we fitted the models using data from only the subjects with all variables recorded.

# References

1. Straczkiewicz, M., James, P. & Onnela, J.-P. A systematic review of smartphone-based human activity recognition methods for health research. *npj Digit. Med.* **4**, 148 (2021).

2. Karas, M. *et al.* Accelerometry Data in Health Research: Challenges and Opportunities: Review and Examples. *Stat. Biosci.* **11**, (2019).

3. Migueles, J. H. *et al.* Calibration and Cross-Validation of Accelerometer Cut-Points to Classify Sedentary Time and Physical Activity from Hip and Non-Dominant and Dominant Wrists in Older Adults. *Sensors (Basel).* **21**, (2021).

4. Migueles, J. H. *et al.* Accelerometer Data Collection and Processing Criteria to Assess Physical Activity and Other Outcomes: A Systematic Review and Practical Considerations. *Sports Med.* **47**, 1821–1845 (2017).

5. Montoye, A. H. K. *et al.* Development of cut-points for determining activity intensity from a wrist-worn ActiGraph accelerometer in free-living adults. *J. Sports Sci.* **38**, 2569–2578 (2020).

6. Jenni, O. G., Chaouch, A., Caflisch, J. & Rousson, V. Infant motor milestones: poor predictive value for outcome of healthy children. *Acta Paediatr.* **102**, e181-4 (2013).

7. Williams, P. T. & Thompson, P. D. Walking versus running for hypertension, cholesterol, and diabetes mellitus risk reduction. *Arterioscler. Thromb. Vasc. Biol.* **33**, 1085–1091 (2013).

8. Hanson, S. & Jones, A. Is there evidence that walking groups have health benefits? A systematic review and meta-analysis. *Br. J. Sports Med.* **49**, 710 LP – 715 (2015).

9. Yaffe, K., Barnes, D., Nevitt, M., Lui, L. Y. & Covinsky, K. A prospective study of physical activity and cognitive decline in elderly women: women who walk. *Arch. Intern. Med.* **161**, 1703–1708 (2001).

10. Pereira, M. A. *et al.* A Randomized Walking Trial in Postmenopausal Women: Effects on Physical Activity and Health 10 Years Later. *Arch. Intern. Med.* **158**, 1695–1701 (1998).

11. Jefferis, B. J., Whincup, P. H., Papacosta, O. & Wannamethee, S. G. Protective effect of time spent walking

on risk of stroke in older men. *Stroke* **45**, 194–199 (2014).

12. Ray, E. L., Sasaki, J. E., Freedson, P. S. & Staudenmayer, J. Physical activity classification with dynamic discriminative methods. *Biometrics* **74**, 1502–1511 (2018).

13. Hills, A. P. & Parker, A. W. Gait characteristics of obese children. *Arch. Phys. Med. Rehabil.* **72**, 403–407 (1991).

14. Balasubramanian, C. K., Neptune, R. R. & Kautz, S. A. Variability in spatiotemporal step characteristics and its relationship to walking performance post-stroke. *Gait Posture* **29**, 408–414 (2009).

15. Urbanek, J. K. *et al.* Validation of Gait Characteristics Extracted From Raw Accelerometry During Walking Against Measures of Physical Function, Mobility, Fatigability, and Fitness. *J. Gerontol. A. Biol. Sci. Med. Sci.* **73**, 676–681 (2018).

16. Del Rosario, M. B. *et al.* A comparison of activity classification in younger and older cohorts using a smartphone. *Physiol. Meas.* **35**, 2269–2286 (2014).

17. Albert, M. V, Toledo, S., Shapiro, M. & Kording, K. Using mobile phones for activity recognition in Parkinson's patients. *Front. Neurol.* **3**, 158 (2012).

18. The method proposed in this paper is available as the Oak tree within the Forest library: https://github.com/onnela-lab/forest.

19. Ellis, K., Kerr, J., Godbole, S., Staudenmayer, J. & Lanckriet, G. Hip and Wrist Accelerometer Algorithms for Free-Living Behavior Classification. *Med. Sci. Sports Exerc.* **48**, 933–940 (2016).

20. Hickey, A., Del Din, S., Rochester, L. & Godfrey, A. Detecting free-living steps and walking bouts: validating an algorithm for macro gait analysis. *Physiol. Meas.* **38**, N1–N15 (2017).

21. Onnela, J.-P. Opportunities and challenges in the collection and analysis of digital phenotyping data. *Neuropsychopharmacology* **46**, 45–54 (2021).

22. Yurur, O., Labrador, M. & Moreno, W. Adaptive and energy efficient context representation framework in mobile sensing. *IEEE Trans. Mob. Comput.* **13**, 1681–1693 (2014).

23. Davis, J. J., Straczkiewicz, M., Harezlak, J. & Gruber, A. H. CARL: a running recognition algorithm for free-living accelerometer data. *Physiol. Meas.* **42**, 115001 (2021).

24. Gjoreski, H. *et al.* The University of Sussex-Huawei Locomotion and Transportation Dataset for Multimodal Analytics with Mobile Devices. *IEEE Access* **6**, 42592–42604 (2018).

25. Esmaeili Kelishomi, A., Garmabaki, A. H. S., Bahaghighat, M. & Dong, J. Mobile User Indoor-Outdoor Detection Through Physical Daily Activities. *Sensors (Switzerland)* **19**, 511 (2019).

26. Müller, J., Müller, S., Baur, H. & Mayer, F. Intra-individual gait speed variability in healthy children aged 1-15 years. *Gait Posture* **38**, 631–636 (2013).

27. Peel, N. M., Kuys, S. S. & Klein, K. Gait Speed as a Measure in Geriatric Assessment in Clinical Settings: A Systematic Review. *Journals Gerontol. Ser. A* **68**, 39–46 (2013).

28. Straczkiewicz, M., Urbanek, J. K., Fadel, W. F., Crainiceanu, C. M. & Harezlak, J. Automatic car driving detection using raw accelerometry data. *Physiol. Meas.* **37**, 1757–1769 (2016).

29. Gjoreski, M. *et al.* Classical and deep learning methods for recognizing human activities and modes of transportation with smartphone sensors. *Inf. Fusion* **62**, 47–62 (2020).

30. Murray, M. P. Gait as a total pattern of movement. *Am. J. Phys. Med.* **46**, 290–333 (1967).

31. Sztyler, T. & Stuckenschmidt, H. On-body localization of wearable devices: An investigation of position-aware activity recognition. in *2016 IEEE International Conference on Pervasive Computing and Communications (PerCom)* 1–9 (2016). doi:10.1109/PERCOM.2016.7456521.

32. Pachi, A. & Ji, T. Frequency and velocity of people walking. *Struct. Eng.* **84**, 36–40 (2005).

33. BenAbdelkader, C., Cutler, R. & Davis, L. Stride and cadence as a biometric in automatic person identification and verification. in *Proceedings of Fifth IEEE International Conference on Automatic Face Gesture Recognition* 372–377 (2002). doi:10.1109/AFGR.2002.1004182.

34. Scholz, R. *The Technique Of The Violin*. (Kessinger Publishing, LLC, 1900).

35. Hagedorn, P. & DasGupta, A. Appendix B: Harmonic Waves and Dispersion Relation. in *Vibrations and*

*Waves in Continuous Mechanical Systems* 367–372 (John Wiley & Sons, Ltd, 2007). doi:https://doi.org/10.1002/9780470518434.app2.

36. Olhede, S. C. & Walden, A. T. Generalized Morse wavelets. *IEEE Trans. Signal Process.* **50**, 2661–2670 (2002).

37. Lilly, J. M. & Olhede, S. C. Higher-Order Properties of Analytic Wavelets. *IEEE Trans. Signal Process.* **57**, 146–160 (2009).

38. Lilly, J. M. jLab: A data analysis package for Matlab, v 1.6.6. http://www.jmlilly.net/jmlsoft.html (2019).

39. Straczkiewicz, M., Glynn, N. W. & Harezlak, J. On Placement, Location and Orientation of Wrist-Worn Tri-Axial Accelerometers during Free-Living Measurements. *Sensors (Basel).* **19**, (2019).

40. Lockhart, J. W. *et al.* Design Considerations for the WISDM Smart Phone-Based Sensor Mining Architecture. in *Proceedings of the Fifth International Workshop on Knowledge Discovery from Sensor Data* 25–33 (Association for Computing Machinery, 2011). doi:10.1145/2003653.2003656.

41. Shoaib, M., Bosch, S., Incel, O. D., Scholten, H. & Havinga, P. J. M. Complex Human Activity Recognition Using Smartphone and Wrist-Worn Motion Sensors. *Sensors* **16**, (2016).

42. Leutheuser, H., Schuldhaus, D. & Eskofier, B. M. Hierarchical, Multi-Sensor Based Classification of Daily Life Activities: Comparison with State-of-the-Art Algorithms Using a Benchmark Dataset. *PLoS One* **8**, 1–11 (2013).

43. Vaizman, Y., Ellis, K. & Lanckriet, G. Recognizing Detailed Human Context in the Wild from Smartphones and Smartwatches. *IEEE Pervasive Comput.* **16**, 62–74 (2017).

44. Anguita, D., Ghio, A., Oneto, L., Parra, X. & Reyes-Ortiz, J. L. A Public Domain Dataset for Human Activity Recognition using Smartphones. in *ESANN* (2013).

45. Ichino, H., Kaji, K., Sakurada, K., Hiroi, K. & Kawaguchi, N. HASC-PAC2016: Large Scale Human Pedestrian Activity Corpus and Its Baseline Recognition. in *Proceedings of the 2016 ACM International Joint Conference on Pervasive and Ubiquitous Computing: Adjunct* 705–714 (Association for Computing Machinery, 2016).

46. Bruno, B., Mastrogiovanni, F., Sgorbissa, A., Vernazza, T. & Zaccaria, R. Analysis of human behavior recognition algorithms based on acceleration data. *2013 IEEE Int. Conf. Robot. Autom.* 1602–1607 (2013).

47. Karas, M. *et al.* Adaptive empirical pattern transformation (ADEPT) with application to walking stride segmentation. *Biostatistics* **22**, 331–347 (2019).

48. Baños, O. *et al.* mHealthDroid: A Novel Framework for Agile Development of Mobile Health Applications. in *IWAAL* (2014).

49. Vavoulas., G., Chatzaki., C., Malliotakis., T., Pediaditis., M. & Tsiknakis., M. The MobiAct Dataset: Recognition of Activities of Daily Living using Smartphones. in *Proceedings of the International Conference on Information and Communication Technologies for Ageing Well and e-Health - Volume 1: ICT4AWE, (ICT4AGEINGWELL 2016)* 143–151 (SciTePress, 2016). doi:10.5220/0005792401430151.

50. Malekzadeh, M., Clegg, R. G., Cavallaro, A. & Haddadi, H. Mobile Sensor Data Anonymization. in *Proceedings of the International Conference on Internet of Things Design and Implementation* 49–58 (ACM, 2019). doi:10.1145/3302505.3310068.

51. Shoaib, M., Bosch, S., Durmaz Incel, O., Scholten, H. & Havinga, P. J. M. Fusion of smartphone motion sensors for physical activity recognition. *Sensors (Switzerland)* **14**, 10146–10176 (2014).

52. Mattfeld, R., Jesch, E. & Hoover, A. A new dataset for evaluating pedometer performance. in *2017 IEEE International Conference on Bioinformatics and Biomedicine (BIBM)* 865–869 (2017). doi:10.1109/BIBM.2017.8217769.

53. Jain, M., Singh, A. P., Bali, S. & Kaul, S. Speed-Breaker Early Warning System. in *NSDR* (2012).

54. Özdemir, A. T. & Barshan, B. Detecting Falls with Wearable Sensors Using Machine Learning Techniques. *Sensors* **14**, 10691–10708 (2014).

55. Sucerquia, A., López, J. D. & Vargas-Bonilla, J. F. SisFall: A Fall and Movement Dataset. *Sensors (Basel).* **17**, (2017).

56. John, D., Tang, Q., Albinali, F. & Intille, S. An Open-Source Monitor-Independent Movement Summary for

Accelerometer Data Processing. *J. Meas. Phys. Behav.* **2**, 268–281.

57. Micucci, D., Mobilio, M. & Napoletano, P. UniMiB SHAR: A dataset for human activity recognition using acceleration data from smartphones. *Appl. Sci.* **7**, 1101 (2017).

58. Weiss, G. M., Yoneda, K. & Hayajneh, T. Smartphone and Smartwatch-Based Biometrics Using Activities of Daily Living. *IEEE Access* **7**, 133190–133202 (2019).

59. Huang, E. J. & Onnela, J.-P. Augmented Movelet Method for Activity Classification Using Smartphone Gyroscope and Accelerometer Data. *Sensors* **20**, (2020).

60. Kang, X., Huang, B. & Qi, G. A Novel Walking Detection and Step Counting Algorithm Using Unconstrained Smartphones. *Sensors (Basel).* **18**, (2018).

61. Casado, F. E. *et al.* Walking Recognition in Mobile Devices. *Sensors* vol. 20 (2020).

62. Huan, Z., Chen, X., Lv, S. & Geng, H. Gait Recognition of Acceleration Sensor for Smart Phone Based on Multiple Classifier Fusion. *Math. Probl. Eng.* **2019**, 6471532 (2019).

63. Cola, G., Avvenuti, M., Musso, F. & Vecchio, A. Personalized gait detection using a wrist-worn accelerometer. in *2017 IEEE 14th International Conference on Wearable and Implantable Body Sensor Networks (BSN)* 173–177 (2017). doi:10.1109/BSN.2017.7936035.

64. Gjoreski, M., Gjoreski, H., Lustrek, M. & Gams, M. How Accurately Can Your Wrist Device Recognize Daily Activities and Detect Falls? *Sensors (Basel).* **16**, (2016).

65. Dijkstra, B., Zijlstra, W., Scherder, E. & Kamsma, Y. Detection of walking periods and number of steps in older adults and patients with Parkinson's disease: accuracy of a pedometer and an accelerometry-based method. *Age Ageing* **37**, 436–441 (2008).

66. Urbanek, J. K. *et al.* Prediction of sustained harmonic walking in the free-living environment using raw accelerometry data. *Physiol. Meas.* **39**, 02NT02 (2018).

67. Mannini, A., Rosenberger, M., Haskell, W. L., Sabatini, A. M. & Intille, S. S. Activity Recognition in Youth Using Single Accelerometer Placed at Wrist or Ankle. *Med. Sci. Sports Exerc.* **49**, 801–812 (2017).

68. Sasaki, J. E. *et al.* Performance of Activity Classification Algorithms in Free-Living Older Adults. *Med. Sci. Sports Exerc.* **48**, 941–950 (2016).

69. Xiao, L. *et al.* Movement prediction using accelerometers in a human population. *Biometrics* **72**, 513–524 (2016).

70. Zhang, S., Rowlands, A. V, Murray, P. & Hurst, T. L. Physical activity classification using the GENEA wrist-worn accelerometer. *Med. Sci. Sports Exerc.* **44**, 742–748 (2012).

**Table 1. Summary of datasets included in this study**. Age, height, weight, and BMI are provided as range (mean±SD), when available.

| Dataset name | Dataset acronym | Population | | | | | | Investigated activities | Measurement parameters | | | | | | Ref. |
|---|---|---|---|---|---|---|---|---|---|---|---|---|---|---|---|
| | | N | Gender (male) | Age (y) | Height (cm) | Weight (kg) | BMI (kg/m²) | | Condition | Sensing device | Approximate sensor location | MR (g) | AR (bit) | SR (Hz) | |
| WISDMs Actitracker activity prediction dataset v2.0 | Actitracker | 166* | 15 (6 females) | 19-51 (30.5±10.8) | 163-188 (174.5±6.8) | 51-109 (75.8±16.6) | 19-35 (24.8±4.6) | Normal walking, sitting, standing, lying, running | Free-living | Smartphone: Android-based | Unspecified | N/A | N/A | 20 | 40 |
| Complex Human Activity Dataset | CHA | 10** | 10 | 23-35 | N/A | N/A | N/A | Normal walking, ascending stairs, descending stairs, sitting, standing, typing, handwriting, eating, drinking, jogging, cycling, giving a talk, smoking | Controlled | Smartphone: Samsung Galaxy S2 | Thigh and wrist | N/A | N/A | 50 | 41 |
| Daily Life Activities Dataset | DaLiAc | 19 | 11 | 18-55 (26.5±7.7) | 158-196 (177.0±11.1) | 54-108 (75.2±14.2) | 17-34 (23.9±3.7) | Normal walking, ascending stairs, descending stairs, lying, sitting, standing, washing dishes, vacuuming, sweeping, running, cycling (50W and 100W), rope jumping | Controlled | Wearable accelerometer: SHIMMER | Waist, chest, and wrist | ±6 | 12 | 204.8 | 42 |
| Dataset for behavioral context recognition in-the-wild from mobile sensor | Extrasensory | 60** | 26 | 18-42 (24±5) | 145-188 (171±9) | 50-93 (66±11) | 18-32 (23±3) | Normal walking, ascending stairs, descending stairs, sitting, standing, lying, watching TV, handwriting, eating, using motorized transportation (car, bus, motor, train), cooking, washing dishes, dressing, grooming, sweeping, running, cycling, jumping, skateboarding | Free-living | Smartphone: Android- and iOS-based | Unspecified | N/A | N/A | 33 | 43 |
| Human Activities and Postural Transitions Dataset | HAPT | 30** | N/A | 19-48 | N/A | N/A | N/A | Normal walking, ascending stairs, descending stairs, standing, sitting, lying, body transitions (standing to sitting, sitting to standing, sitting to lying, lying to sitting, standing to lying, and lying to standing) | Controlled | Smartphone: Samsung Galaxy S2 | Waist | N/A | N/A | 50 | 44 |
| Human Activity Sensing Consortium – Pedestrian Activity Corpus 2016 | HASC | 539 | 438 | 15-69 (28.6±12.2) | 147-189 (169.4±7.9) | 37-118 (62.8±11.5) | 15-38 (21.8±3.4) | Normal walking, ascending stairs, descending stairs, standing, jogging, jumping | Controlled, Free-living | Smartphone: Android- and iOS-based | Thigh, waist, chest, arm, wrist, and unspecified | N/A | N/A | 100 | 45 |
| Public Dataset of Accelerometer Data for Human Motion Primitives Detection | HMPD | 16** | 11 | 19-81 (57.4) | N/A | 56-85 (72.7) | N/A | Normal walking, ascending stairs, descending stairs, drinking, pouring, eating soup or meat, combing hair, brushing teeth, using telephone, body transitions (standing to lying, lying to standing, standing to sitting, and sitting to standing) | Controlled | Wearable accelerometer | Wrist | ±1.5 | 6 | 32 | 46 |
| Identification of Walking, Stair Climbing, and Driving using Wearable Accelerometers | IWSCD | 32 | 13 | 23-52 (39.0±9.0) | 147-193 (173.5±11.1) | 45-140 (77.0±22.9) | 18-40 (25.2±5.6) | Normal walking, ascending stairs, descending stairs, using motorized transportation (car) | Controlled | Wearable accelerometer: ActiGraph GT3X+ | Waist and wrist | N/A | N/A | 100 | 47 |
| Mobile Health Dataset | MHEALTH | 10** | N/A | N/A | N/A | N/A | N/A | Normal walking, lying, sitting, standing, cycling, jogging, running, forward and backward jumping, body stretching (bending waist forward, elevating arm, crouching) | Controlled | Wearable accelerometer: SHIMMER2 | Chest and wrist | N/A | N/A | 50 | 48 |
| Recognition of Activities of Daily Living using Smartphones | MobiAct | 61 | 42 | 20-40 (24.9±3.7) | 158-193 (175.9±8.1) | 50-120 (76.8±15.0) | 18-35 (24.7±3.8) | Normal walking, ascending stairs, descending stairs, lying, sitting, standing, jogging, jumping, body transitions (standing to sitting [on a chair, in a car], sitting to standing [from a chair, from a car]) | Controlled | Smartphone: Samsung Galaxy S3 | Thigh | N/A | N/A | 200 | 49 |
| Sensor Based Human Activity and Attribute Recognition | MotionSense | 24 | 14 | 18-46 (28.8±5.4) | 161-190 (174.2±8.9) | 48-102 (72.1±16.2) | 18-32 (23.6±4.1) | Normal walking, ascending stairs, descending stairs, sitting, standing, jogging | Controlled | Smartphone: iPhone 6 | Thigh | N/A | N/A | 50 | 50 |
| Physical Activity Recognition Dataset Using Smartphone Sensors | PARDUSS | 10** | 10 | 25-30 | N/A | N/A | N/A | Normal walking, ascending stairs, descending stairs, sitting, standing, jogging, cycling | Controlled | Smartphone: Samsung Galaxy S2 | Thigh, waist, arm, and wrist | N/A | N/A | 50 | 51 |
| Pedometer Evaluation Project | Pedometer | 30 | 15 | 19-27 (21.9±52.4) | 152-193 (171.0±10.8) | 43-136 (70.5±17.6) | 17-37 (23.8±3.7) | Normal walking | Controlled | Wearable sensor: SHIMMER3 | Waist and wrist | ±4 | N/A | 15 | 52 |
| Real-World Dataset | RealWorld | 15 | 8 | 16-62 (31.9±12.4) | 163-183 (173.1±6.9) | 48-95 (74.1±13.8) | 18-35 (24.7±4.4) | Normal walking, ascending stairs, descending stairs, lying, sitting, standing, running, jumping | Controlled | Smartphone: Samsung Galaxy S4, smartwatch: LG G Watch R | Smartphone: thigh, waist, chest, and arm, smartwatch: wrist | N/A | N/A | 50 | 31 |
| Speed-Breaker Dataset | SpeedBreaker | 40** | N/A | N/A | N/A | N/A | N/A | Using motorized transportation (car, motorcycle, and rickshaw) | Free-living | Smartphone: Android-based | Unspecified | N/A | N/A | 100 | 53 |
| Simulated Falls and Daily Living Activities Dataset | SFDLA | 17 | 10 | 19-27 (21.9±2.0) | 157-184 (171.6±7.8) | 47-92 (65.0±13.9) | 17-31 (21.9±3.7) | Normal walking, walking backwards, limping, jogging, squatting, bending, body transitions (lying to sitting, lying to standing, and standing to sitting [on a chair, a sofa, a bed, in the air], coughing/sneezing | Controlled | Wearable accelerometer: Xsens MTw | Thigh, waist, chest, and wrist | ±12 | N/A | 25 | 54 |
| A Fall and Movement Dataset | SisFall | 38 | 19 | 19-75 (40.2±21.3) | 149-183 (164.1±9.3) | 41-102 (62.2±12.6) | 18-35 (23.0±3.5) | Normal walking (slow, fast), jogging (slow and fast), jumping, body transitions (rolling while lying, standing to sitting to standing [with a low and a high chair, in a car, slow and fast], sitting to lying to sitting [slow and fast], and sitting to standing to sitting) | Controlled | Wearable accelerometer: self-developed | Waist | ±16 | 13 | 200 | 55 |
| Human physical activity dataset | SPADES | 42 | 27 | 18-30 (23±3) | 151-180 174±8 | 51-112 (73±15) | 18-35 (24±4) | Normal walking, ascending stairs, descending stairs, treadmill walk (1, 2, 3, and 3.5 mph), lying, sitting, standing, reclining, | Controlled | Wearable accelerometer: ActiGraph GT9X | Thigh, waist, and wrist | 8 | N/A | 80 | 56 |

| | | | | | | handwriting, typing, folding towels, filling shelves, sweeping, running, cycling, jumping jacks | | | | | | |
|---|---|---|---|---|---|---|---|---|---|---|---|---|
| University of Milano Bicocca Smartphone-based Human Activity Recognition Dataset | UniMiB-SHAR | 30 | 6 | 18-60 (26.6±11.6) | 160-190 (168.8±6.8) | 50-82 (64.4±9.8) | 18-27 (22.5±2.5) | Normal walking, ascending stairs, descending stairs, running, jumping, body transitions (lying to standing, sitting to standing, standing to sitting, and standing to lying) | Controlled | Smartphone: Samsung Galaxy Nexus | Thigh | ±2 | 9 | 50 | [57] |
| Wireless Sensor Data Mining Dataset | WISDM | 51** | N/A | 18-25 | N/A | N/A | N/A | Normal walking, sitting, standing, jogging, eating soup, pasta, and chips, drinking, handwriting, typing, folding clothes, brushing teeth, clapping | Controlled | Smartphone: Google Nexus 5/5X and Samsung Galaxy S5, smartwatch: LG watch | Smartphone: thigh, smartwatch: wrist | N/A | N/A | 20 | [58] |

Note: * detailed demographics available for some subjects; ** detailed demographics unavailable; BMI – body mass index; MR – measurement range; AR – amplitude resolution; SR – approximate sampling rate.

**Table 2**. Walking classification accuracy across all subjects and activities. The accuracy is provided as mean (95% CI), sample size. For walking activities, the metric indicates sensitivity; for non-walking activities, the metric indicates specificity.

| Locations typical to | Smartphone | | | | | Smartwatch |
|---|---|---|---|---|---|---|
| | Thigh | Waist | Chest | Arm | Unspecified | Wrist |
| Walking | | | | | | |
|   Normal walking | 0.92 (0.91,0.93), 459 | 0.95 (0.94,0.97), 538 | 0.97 (0.95,0.98), 110 | 0.92 (0.88,0.96), 60 | 0.93 (0.91,0.94), 273 | 0.92 (0.9,0.94), 352 |
| Stair climbing | | | | | | |
|   Ascending stairs | 0.83 (0.8,0.85), 361 | 0.85 (0.83,0.88), 396 | 0.93 (0.89,0.96), 74 | 0.82 (0.76,0.88), 60 | 0.9 (0.85,0.94), 69 | 0.73 (0.68,0.77), 222 |
|   Descending stairs | 0.85 (0.83,0.87), 364 | 0.86 (0.84,0.89), 392 | 0.78 (0.71,0.85), 74 | 0.81 (0.73,0.89), 62 | 0.83 (0.76,0.89), 70 | 0.73 (0.69,0.77), 213 |
| Treadmill | | | | | | |
|   1 mph | 0.19 (0.13,0.24), 31 | 0.02 (0.01,0.03), 31 | - | - | - | 0 (0,0), 31 |
|   2 mph | 0.82 (0.73,0.91), 30 | 0.77 (0.67,0.87), 30 | - | - | - | 0.05 (0.01,0.09), 30 |
|   3 mph | 0.96 (0.95,0.98), 29 | 0.99 (0.98,1), 29 | - | - | - | 0.91 (0.85,0.97), 29 |
|   3.5 mph | 0.97 (0.94,0.99), 28 | 0.99 (0.98,1), 28 | - | - | - | 0.79 (0.68,0.9), 28 |
|   Other walking | 0.81 (0.72,0.9), 17 | 0.68 (0.52,0.83), 17 | 0.79 (0.69,0.88), 17 | - | - | 0.47 (0.31,0.63), 17 |
| Non-walking | | | | | | |
|   Stationary & TV | 0.99 (0.99,1), 401 | 1 (1,1), 380 | 0.99 (0.99,1), 89 | 1 (0.99,1), 60 | 1 (0.99,1), 133 | 0.99 (0.98,0.99), 257 |
|   Desk work | 1 (1,1), 153 | 1 (1,1), 33 | - | - | 0.99 (0.98,0.99), 35 | 1 (1,1), 154 |
|   Eating | 1 (0.99,1), 212 | - | - | - | 0.97 (0.97,0.98), 57 | 0.99 (0.98,1), 214 |
|   Drinking | 0.99 (0.98,1.01), 61 | - | - | - | - | 1 (0.99,1), 81 |
|   Motorized transport | - | 1 (1,1), 32 | - | - | 0.92 (0.9,0.93), 117 | 1 (0.99,1), 32 |
| Household | | | | | | |
|   Sweeping | 0.94 (0.91,0.97), 34 | 0.95 (0.94,0.97), 53 | 0.9 (0.88,0.93), 19 | - | 0.99 (0.97,1.01), 2 | 0.57 (0.51,0.62), 53 |
|   Vacuuming | - | 0.98 (0.96,0.99), 19 | 1 (0.99,1), 19 | - | - | 0.92 (0.84,1), 19 |
|   Folding clothes | 0.97 (0.95,1), 51 | - | - | - | - | 0.73 (0.68,0.78), 51 |
|   Washing dishes | - | 1 (1,1), 19 | 1 (1,1), 19 | - | 0.99 (0.98,1.01), 5 | 0.96 (0.93,0.99), 19 |
|   Grooming | - | - | - | - | 0.93 (0.86,1.01), 15 | - |
|   Dressing | - | - | - | - | 0.91 (0.75,1.07), 8 | - |
|   Cooking | - | - | - | - | 0.98 (0.97,1), 12 | - |
|   Filling shelves | 0.94 (0.9,0.97), 32 | 0.98 (0.98,0.99), 32 | - | - | - | 0.74 (0.69,0.79), 32 |
| Personal hygiene | | | | | | |
|   Combing hair | - | - | - | - | - | 0.67 (0.44,0.9), 5 |
|   Brushing teeth | 1 (1,1), 51 | - | - | - | - | 0.98 (0.97,0.99), 54 |
| Sports | | | | | | |
|   Running | 0.94 (0.93,0.96), 431 | 0.97 (0.96,0.98), 430 | 0.95 (0.92,0.98), 114 | 0.97 (0.95,0.99), 60 | 0.92 (0.89,0.95), 126 | 0.97 (0.96,0.99), 264 |
|   Cycling | 0.96 (0.94,0.98), 62 | 0.97 (0.94,0.99), 90 | 0.99 (0.97,1.01), 48 | 0.99 (0.99,1), 10 | 0.84 (0.75,0.92), 23 | 0.99 (0.98,1), 110 |
|   Jumping | 0.14 (0.11,0.17), 326 | 0.18 (0.15,0.22), 354 | 0.13 (0.07,0.19), 85 | 0.13 (0.06,0.2), 50 | 0.14 (0.06,0.21), 61 | 0.21 (0.17,0.26), 176 |
| Other | | | | | | |
|   Hand clapping | 0.98 (0.96,1), 51 | - | - | - | - | 0.93 (0.9,0.96), 51 |
|   Smoking | 1 (1,1), 10 | - | - | - | - | 1 (1,1), 10 |
|   Giving a talk | 1 (1,1), 10 | - | - | - | - | 0.96 (0.92,1), 10 |
|   Body transitions | 0.97 (0.95,0.98), 108 | 0.99 (0.98,1), 85 | 0.93 (0.89,0.98), 17 | - | - | 1 (0.99,1), 32 |
|   Using a phone | 1 (1,1), 17 | 1 (1,1), 17 | 1 (1,1), 17 | - | - | 1 (1,1), 17 |
|   Coughing | 1 (1,1), 17 | 1 (1,1), 17 | 1 (1,1), 27 | - | - | 0.79 (0.67,0.91), 27 |

**Table 3. Bias model estimates. a.** Coefficient estimates, standard errors, and 95% confidence intervals for *StandardReg* model. **b.** Coefficient estimates, standard errors, and 95% confidence intervals for the fixed effects in *MixedReg* model. In **a.** and **b.**, the covariate age in years was standardized by centering with the mean (28.7 y) and dividing by the standard deviation (12.0 y). BMI was also standardized by centering with the mean (22.9 kg/m$^2$) and dividing by the standard deviation (3.9 kg/m$^2$). Gender, environmental condition, sensor location, and study are incorporated using indicator variables. The reference category for gender is female, the reference category for environment is the controlled setting, the reference category for sensor location is the arm, and the reference category for study is *UniMiBSHAR*.

|  | Estimate | Standard error | 95% Confidence Interval |
|---|---|---|---|
| **a. *StandardReg*** | | | |
| **Intercept** | 0.7402 | 0.0326 | (0.6763, 0.8041) |
| **Age** | 0.0085 | 0.0044 | (-0.0002, 0.0172) |
| **BMI** | 0.0044 | 0.0041 | (-0.0036, 0.0123) |
| **Gender** | | | |
| Male | -0.0002 | 0.0091 | (-0.0180, 0.0176) |
| **Measurement condition** | | | |
| Free-living | -0.0136 | 0.0171 | (-0.0470, 0.0199) |
| **Sensor location** | | | |
| Thigh | 0.0201 | 0.0220 | (-0.0231, 0.0632) |
| Waist | 0.0618 | 0.0221 | (0.0184, 0.1052) |
| Chest | 0.0638 | 0.0253 | (0.0142, 0.1134) |
| Wrist | 0.0261 | 0.0226 | (-0.0182, 0.0704) |
| Unspecified | 0.0255 | 0.0273 | (-0.0280, 0.0790) |
| **Study** | | | |
| Actitracker | 0.1724 | 0.0483 | (0.0776, 0.2673) |
| DaLiAc | 0.2022 | 0.0317 | (0.1400, 0.2645) |
| HASC | 0.1586 | 0.0263 | (0.1071, 0.2101) |
| IWSCD | 0.1383 | 0.0312 | (0.0770, 0.1995) |
| MobiAct | 0.1811 | 0.0299 | (0.1224, 0.2397) |
| MotionSense | 0.2306 | 0.0361 | (0.1597, 0.3014) |
| Pedometer | 0.2075 | 0.0314 | (0.1458, 0.2692) |
| RealWorld | 0.1782 | 0.0299 | (0.1195, 0.2369) |
| SFDLA | 0.1614 | 0.0302 | (0.1021, 0.2207) |
| SisFall | 0.0559 | 0.0344 | (-0.0116, 0.1234) |
| SPADES | 0.1698 | 0.0282 | (0.1145, 0.2251) |
| **b. *MixedReg*** | | | |
| **Intercept** | 0.7698 | 0.0303 | (0.7050, 0.8288) |
| **Age** | 0.0114 | 0.0051 | (0.0017, 0.0210) |
| **BMI** | 0.0027 | 0.0049 | (-0.0082, 0.0111) |
| **Gender** | | | |
| Male | -0.0021 | 0.0109 | (-0.0257, 0.0185) |
| **Measurement condition** | | | |
| Free-living | -0.0139 | 0.0179 | (-0.0509, 0.0222) |
| **Sensor location** | | | |
| **Thigh** | -0.0088 | 0.0181 | (-0.0432, 0.0239) |
| **Waist** | 0.0239 | 0.0178 | (-0.0102, 0.0604) |
| **Chest** | 0.0277 | 0.0205 | (-0.0158, 0.0696) |
| **Wrist** | -0.0051 | 0.0180 | (-0.0416, 0.0276) |
| **Unspecified** | -0.0175 | 0.0230 | (-0.0629, 0.0295) |
| **Study** | | | |
| **Actitracker** | 0.1880 | 0.0481 | (0.0918, 0.2906) |
| **DaLiAc** | 0.2097 | 0.0359 | (0.1493, 0.2815) |
| **HASC** | 0.1648 | 0.0267 | (0.1157, 0.2227) |
| **IWSCD** | 0.1425 | 0.0332 | (0.0763, 0.2169) |
| **MobiAct** | 0.1834 | 0.0304 | (0.1286, 0.2436) |
| **MotionSense** | 0.2312 | 0.0366 | (0.1631, 0.3014) |
| **Pedometer** | 0.2154 | 0.0334 | (0.1508, 0.2837) |
| **RealWorld** | 0.1764 | 0.0366 | (0.1108, 0.2469) |
| **SFDLA** | 0.1676 | 0.0358 | (0.0905, 0.2358) |
| **SisFall** | 0.0624 | 0.0344 | (-0.0048, 0.1336) |
| **SPADES** | 0.1760 | 0.0307 | (0.1164, 0.2416) |

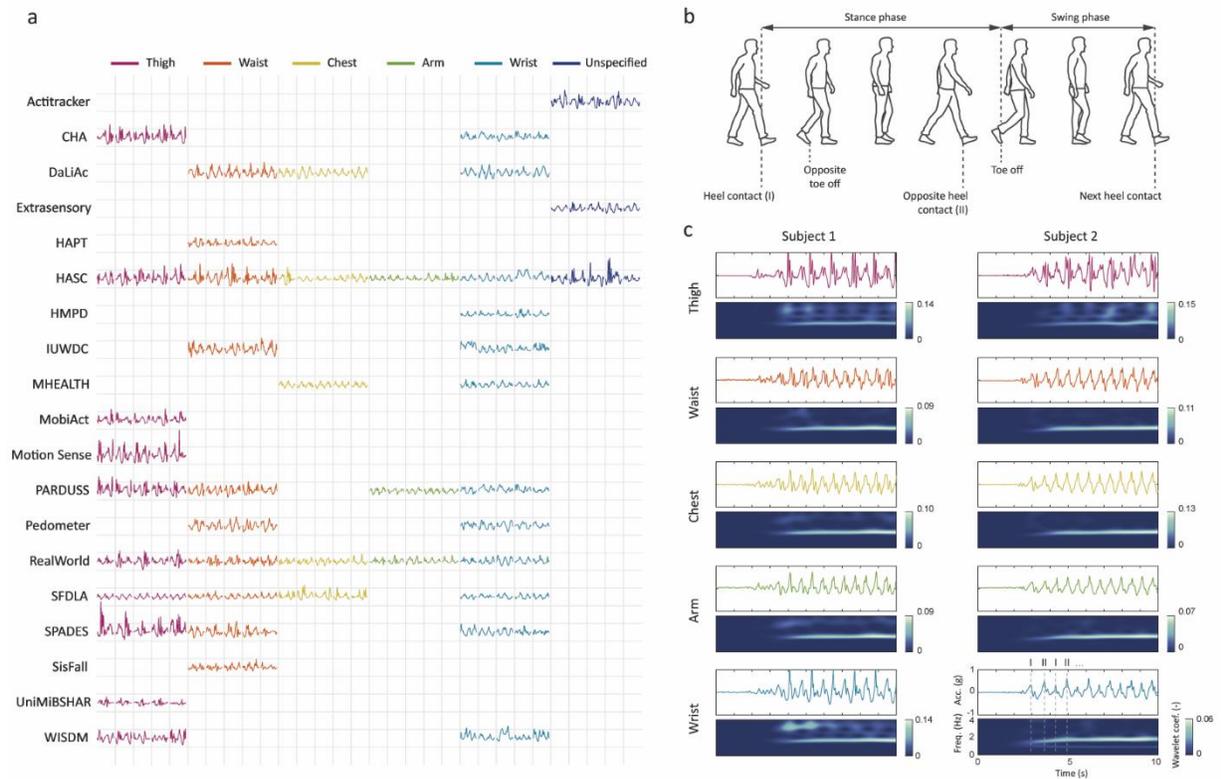

**Figure 1. Human gait and accelerometer data collected using body-worn devices. a.** Vector magnitude of raw accelerometer time series of walking strides measured at different body locations. Strides were extracted for five randomly selected subjects in each study and at each location available in that study. Vertical grid lines separate strides of different subjects, and horizontal grid lines mark stride acceleration equal to +1 g and -1 g above and below, respectively, of the acronym of the corresponding study. Colors indicate approximate locations of sensing devices. **b.** Walking activity is typically understood as a cyclic series of movements initiated the moment the foot contacts the ground, followed by the stance phase (i.e., when the foot is on the ground) and the swing phase (i.e., when the foot is in the air); the cycle is completed when the same foot makes contact with the ground again. **c.** Several examples of resting and walking acceleration signals collected simultaneously using smartphones at different body locations (thigh, waist, chest, arm) and a smartwatch worn on the wrist by two subjects. Corresponding time-frequency representation were computed with CWT.

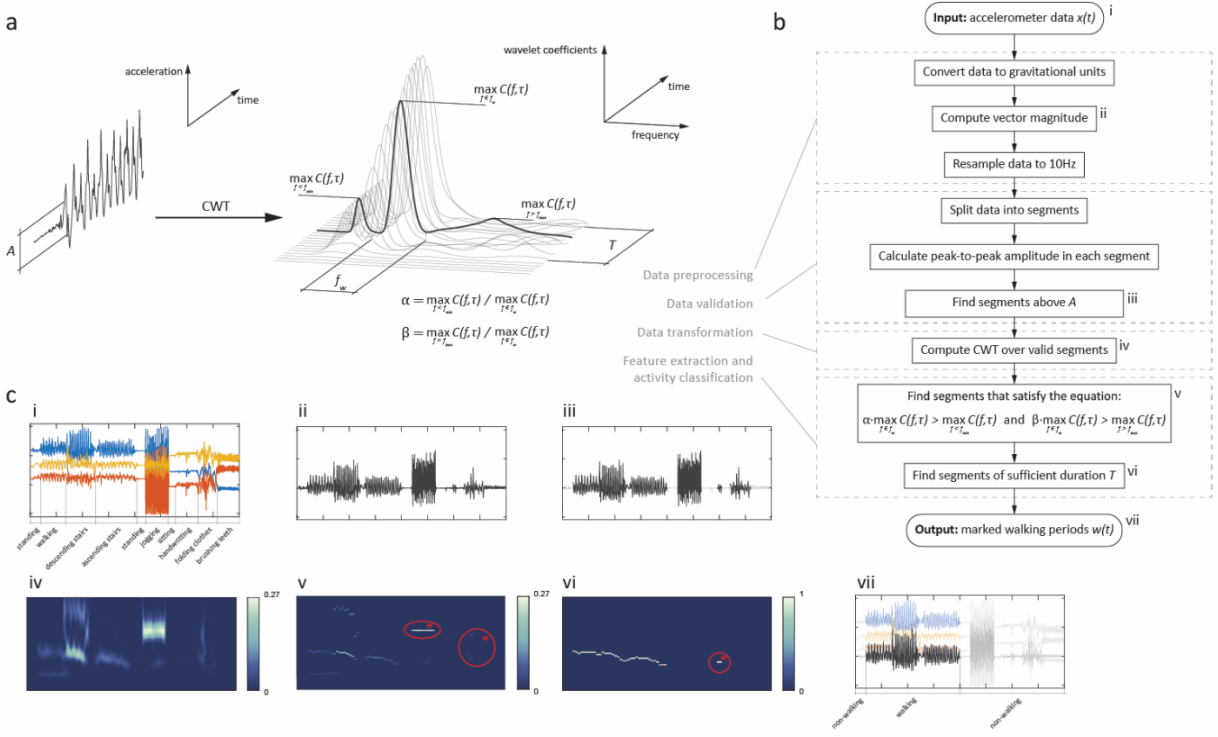

**Figure 2. Walking recognition algorithm and visualization of signal features and data processing steps. a.** Vector magnitudes of raw time-domain accelerometer signal is used to compute peak-to-peak amplitudes in one-second segments, which are then compared to a predefined threshold *A*; segments with amplitude below the threshold are excluded from further processing. Time-frequency decomposition computed using CWT reveals temporal gait features (wavelet coefficients) within, below, and above typical step frequency range $f_w$, used to calculate gait harmonics parameters $\alpha$ and $\beta$. The activity is classified as walking when all amplitude- and frequency-based conditions are satisfied for each least *T* segments (seconds). **b.** Walking recognition algorithm consists of four main blocks: data preprocessing block standardizes the input signal (i) to the common format insensitive to temporal sensor orientation (ii), data validation block finds high-amplitude data segments (iii), data transformation block reveals frequency of temporal oscillations in time (iv), feature extraction and activity classification block excludes segments with important frequency components outside $f_w$ (v), as well as segments of insufficient duration (vi), and returns the output signal with marked walking (vii). **c.** Visualization of walking recognition algorithm steps using example data collected with a smartwatch placed on a wrist (*WISDM* dataset [58]).

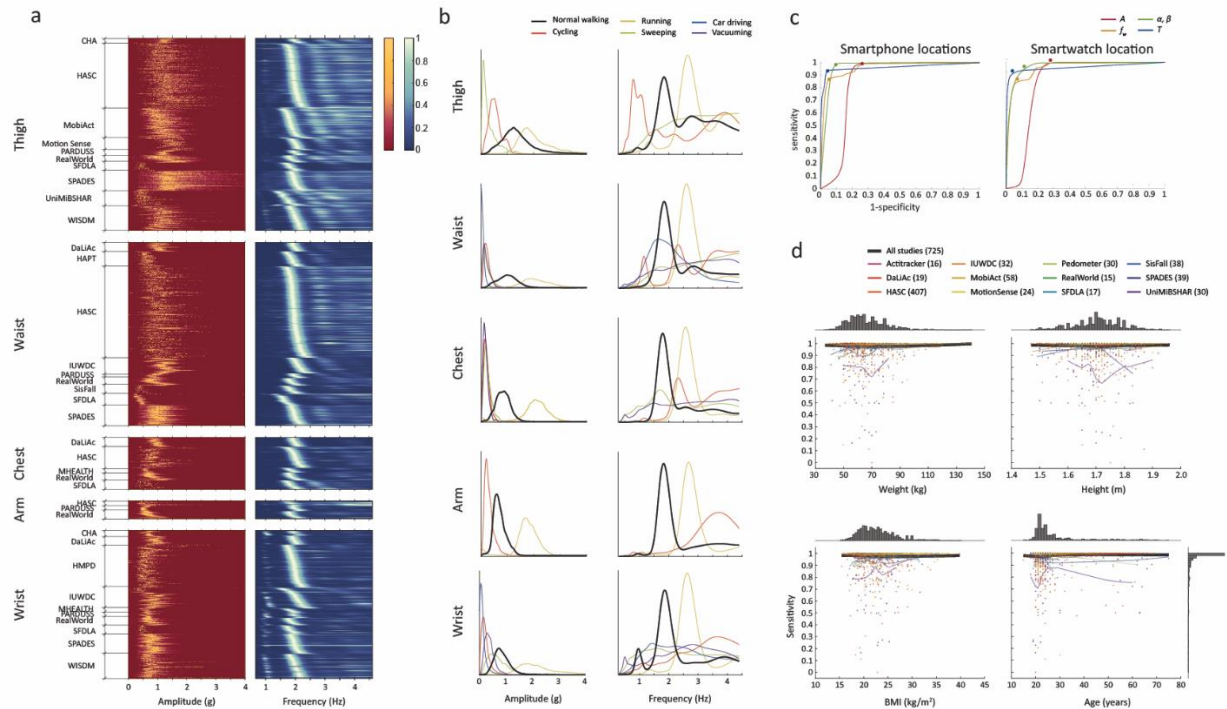

**Figure 3. Exploratory data analysis, tuning parameter selection, and bias assessment. a.** Distribution of accelerometer-based signal features (peak-to-peak amplitude and wavelet coefficients) for various sensor body locations and studies during normal walking. Each row corresponds to a subject while color intensity corresponds to the frequency of a given value for this subject. In each study, subjects were sorted by the location of maximum wavelet coefficient between 1.4 Hz and 2.3 Hz. **b.** Cumulative cross-study distribution of peak-to-peak amplitude and wavelet coefficients for normal walking and other common daily activities for various sensor body locations. Distributions were normalized to have equal area under the curve. Distributions reveal that amplitude- and frequency-based features are well suited to separate walking from other activities. They also reveal visual differences between frequency-based features at locations typical to smartphone (thigh, waist, chest, arm) and smartwatch (wrist). **c.** Receiver-operating characteristics (ROC) used for tuning parameter selection using one vs. all approach (normal walking vs. all non-walking activities). ROCs were computed separately for sensor body locations common to the smartphone and smartwatch. Dots represent optimal cutoff points at which the sum of sensitivity and specificity is maximized. **d.** Normal walking sensitivity metrics against body measure and demographic covariates of weight, height, BMI, and age. Each dot represents a metric for one subject averaged

across body locations available for this subject and activity repetitions this subject performed. The light curves represent smoothed study-level averages while the black curve is an overall average.

**Supplementary Table 1**. Selected methods for walking recognition using body-worn device data (for further read on activity recognition methods using smartphones, see [1]).

| Sensing device | Device location | Sensors involved | Population | Validation setting | Investigated activities | Main concept | Reference |
|---|---|---|---|---|---|---|---|
| Smartphone | Front and back pants pocket | Accelerometer and gyroscope | N=4 (2 males) Age: 27-54 y | Controlled | Walking, ascending stairs, descending stairs, sitting, standing | Tri-axial accelerometer and gyroscope data are used as activity templates. Walking is identified when the distance between new data fragment and activity template exceeds the threshold. | [59] |
| Smartphone | Front pants pocket | Gyroscope | N=8 (5 males), Age: 23-26 y | Controlled | Walking, stair climbing, standing, typing | Fast Fourier transform is computed over most sensitive axis. Walking is identified when the average area under the spectrum within predefined range exceeds area under the spectrum below with range. | [60] |
| Smartphone | Hand, pants pocket, backpack | Accelerometer and gyroscope | N=77 | Controlled | Walking, non-walking | Authors investigate several classification approaches, e.g., using 21 time- and frequency domain features combined with various machine learning techniques, activity-templates, and deep learning. Deep learning provides the highest classification accuracy. | [61] |
| Smartphone | Pants pocket | Accelerometer | N=32 (16 males), age: ~25 y | Controlled | Walking | Tri-axial accelerometer data is filtered using a low-pass filter. Activity templates, time-domain features, and frequency-domain features are extracted for multiple machine learning classifiers. Walking is identified using voting method. | [62] |
| Wearable accelerometer | Wrist | Accelerometer | N=20 (15 males), age: 26.8±3.6 y | Controlled | Normal walking, fast walking, random hand movement | Vector magnitude is segmented using adaptive window based on local maxima that indicate walking steps. Several time-domain features are extracted for each segment. Walking is identified using threshold on anomaly detection score. | [63] |
| Wearable accelerometer | Wrist | Accelerometer | Dataset 1: N=5 (4 males); age: 29.4±2.1 y Dataset 2: N=10 (8 males), age: 27.2±3.1 y Dataset 3: N=3 | Controlled | Dataset 1: walking, cycling, running, lying, sitting, standing, kneeling, bending, body transitions Dataset 2: treadmill walking, treadmill running, stationary bike cycling, sitting, standing, lying, bending, kneeling, body transitions Dataset 3: walking, standing, sitting, lying | Segmented tri-axial data is filtered using low-pass and band-pass filters. Several time-domain features are extracted. Feature selection is performed to identify most informative features. Machine learning (Random Forest) classifier is used to distinguish walking from other activities in the dataset. | [64] |
| Wearable accelerometer | Waist | Accelerometer | Dataset 1: N=20 (10 males); healthy; Age: 68.5±7.4 y Dataset 2: N=32 (17 males); PD; 67.3±6.6 y | Controlled | Normal walking, slow walking, fast walking, walking while carrying a tray | Walking is identified when the absolute resultant of the three axes data exceeds a threshold. | [65] |
| Wearable accelerometer | Hip | Accelerometer | N=49 (25 males) Age: 78 y (IQ: 74-82 y) | Controlled | Fast walking, dressing, shopping, chair stands | Raw tri-axial data are transformed into vector magnitude; vector magnitude is transformed into frequency-domain using Fast Fourier transform; walking is identified when the highest ratio between (a) and (b) exceeds the threshold, where (a) area under the spectrum at baseline frequency and its harmonics for each baseline frequency bin, and (b) is area under the spectrum for all frequencies. | [66] |
| Wearable accelerometer | Lower back | Accelerometer | N=10 (20-33), age: 27.5±4.7 y, healthy | Free-living | Walking in various context, other non-walking activities | Tri-axial data are filtered out from high-frequency noise (17 Hz and above) and transformed to fixed (horizontal-vertical) coordinate system. A moving standard deviation and vertical acceleration detects upright movement. Micro-gait events (initial contact & final contact) detected using continuous wavelet transform are used to identify walking. | [20] |
| Wearable accelerometer | Wrist, ankle | Accelerometer | (1) N=33 (11 males), age: 18-75 y (2) N=20 (12 males), age: 13±1.3 y | Controlled | Various types of walking (while carrying item, on treadmill, etc.), stair climbing, cycling, playing basketball, tennis, and soccer, painting, house cleaning, sitting, lying, sitting | Vector magnitude is preprocessed using low-pass filter (15 Hz). Segmented data are used to extract several time- and frequency-domain. New features based on signal fragmentation are used. Machine learning (SVM) technique is used to distinguish between ambulation, cycling, sedentary, and activities outside of these categories. | [67] |
| Wearable accelerometer | Hip, wrist | Accelerometer | N=40 (0 males), age: 55.2±15.3 y, BMI: 32.0±3.7 | Free-living | Walking/running, sitting, standing, riding in a vehicle | Raw (unfiltered) tri-axial acceleration and its vector magnitude data are used to extract 41 time- and frequency-domain features. Feature vectors are used in Random Forest classifier to predict performed activities. Hidden Markov model is used to smooth predictions over time. | [19] |
| Wearable accelerometer | Wrist, ankle | Accelerometer | Dataset 1: N=33 (11 males), age: 18-75 y Dataset 2: N=35 (14 males), age: 65-80 y | Controlled (1-2), free-living (3) | Ambulation, cycling, sedentary, and activities outside of these categories | Tri-axial data are used to compute 13 or 77 time- and frequency-domain features. Conditional Random Fields is used for recognition of ambulation. | [12] |
| | | | Dataset 3: N=15 (6 males), age: 65-78 y | | | | |
| Wearable accelerometer | Ankle, hip, wrist | Accelerometer | N=35 (14 males), age: 70.8±4.9 y | Free-living | Locomotion (slow walk, self-paced 400m walk, carrying groceries), household (dusting, gardening, vacuuming, self-care, laundry, organizing the room), sedentary (lying down, sitting, crossword puzzles, playing cards), standing (stationary), recreational activities (tai chi, simulated bowling) | Tri-axial data is used to extract several time- and frequency-domain features. Support vector machines and Random Forest methods were used to classify locomotion activities. | 68 |
| Wearable accelerometer | Hip | Accelerometer | N=20, older | Controlled | Normal walking, fast walking, dealing cards, standing from a chair, shopping, handwriting, vacuuming, folding towels, dressing, kneading, washing dishes, standing, lying. | Tri-axial data is transformed into vector magnitude and normalized using linear transformation. Extracted activity templates are used for activity classification based on the minimum distance from an activity pattern. | 69 |
| Wearable accelerometer | Wrist, hip | Accelerometer | N=60 (23 males), age: 40-65 y | Controlled | Lying, standing, seated computer work, treadmill walk, ascending and descending stairs, normal walk, washing windows, washing up, shelf stacking, sweeping, running, | Segmented data is transformed into vector magnitude and used to extract several time- and frequency-domain features. Support vector machines is used to select most informative features. Decision tree is used to distinguish between sedentary, household, walking, and running activities. | 70 |

**Supplementary Table 2.** Activity groups used in our study. Note that activity names in column "Included activities" were given by researchers collecting datasets.

| Activity group | Included activities |
| --- | --- |
| Walking | |
|   Normal walking | Walking, fast walking, slow walking |
| Non-walking | |
|   Stationary | Lying, sitting, standing, standing in the elevator |
|   Desk work & TV | Watching TV, working on the computer, typing, handwriting |
|   Eating | Eating, eating pasta, eating sandwich, eating soup, eating meat |
|   Drinking | Pouring, drinking, drinking coffee |
|   Motorized | Car driving, travelling in the car, motorcycle, and rickshaw |
|   Other | Moving in the elevator, smoking, giving a talk, using telephone, coughing, sneezing, body transitions, clapping |

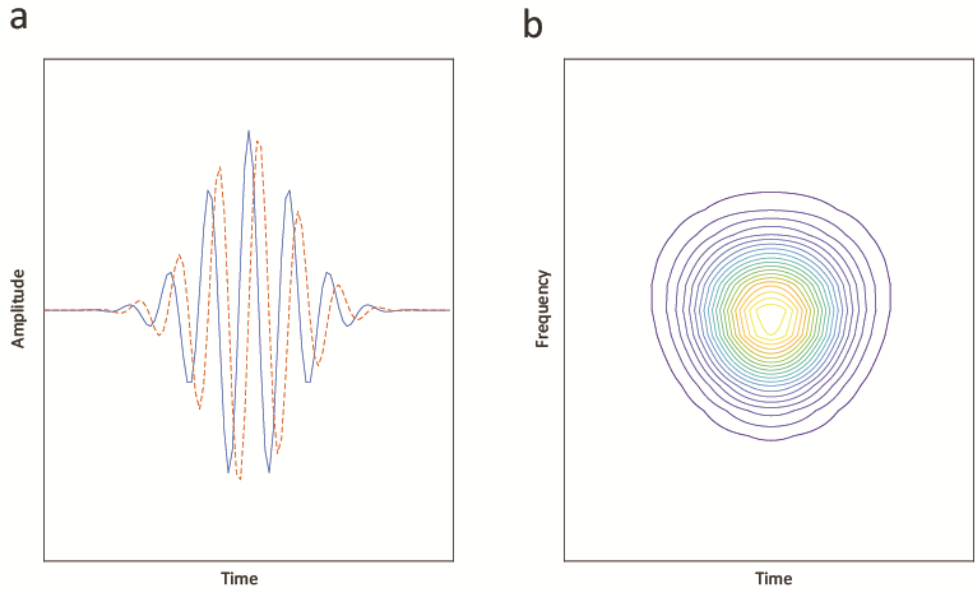

**Supplementary Figure 1. A Mother wavelet, Morse, used in the walking recognition method.** a. The wavelet's real part is displayed with the blue solid line; the imaginary part is displayed with the red dashed line. b. The wavelet has sampling frequency equal to 10 Hz and central frequency equal to 1 Hz. Its symmetry coefficient, $\gamma$, is equal to 3, while time-bandwidth product, $P^2$, is equal to 60, which provide symmetrical coefficients' spread in time and frequency domains.